\documentclass[aps,prx,twocolumn,amsmath,amssymb,superscriptaddress,floatfix]{revtex4-2}

\usepackage{mathtools}
\usepackage{multirow}
\usepackage{xcolor}
\usepackage{bm}
\usepackage{amsfonts,amssymb,amsmath}
\usepackage{graphicx,dcolumn,bm,xcolor,braket,slashed}
\usepackage{times} 
\usepackage{comment}
\usepackage{array}
\usepackage{textcomp}
\usepackage[normalem]{ulem}
\usepackage{dsfont}

\usepackage[title,toc,titletoc,page]{appendix}

\renewcommand{\v}[1]{{\bf #1}}
\newcommand{\be}{\begin{eqnarray}}
\newcommand{\ee}{\end{eqnarray}}
\newcommand{\bbm}{\begin{bmatrix}}
\newcommand{\ebm}{\end{bmatrix}}
\newcommand{\bpm}{\begin{pmatrix}}
\newcommand{\epm}{\end{pmatrix}}

\begin{document}

%\title{Geometry-induced magnetic phase transition}
\title{Magnetic phase transitions driven by quantum geometry}

\author{Chang-geun Oh}
\email{cg.oh.0404@gmail.com}
\affiliation{Department of Applied Physics, The University of Tokyo, Tokyo 113-8656, Japan}
\author{Taisei Kitamura}
%\email{taisei.kitamura@riken.jp}
\affiliation{Department of Physics, Graduate School of Science, Kyoto University, Kyoto 606-8502, Japan}
\affiliation{RIKEN Center for Emergent Matter Science (CEMS), Wako 351-0198, Japan}
\author{Akito Daido}
\affiliation{Department of Physics, Graduate School of Science, Kyoto University, Kyoto 606-8502, Japan}
\author{Jun-Won Rhim}
\affiliation{Department of Physics, Ajou University, Suwon 16499, Republic of Korea}
\author{Youichi Yanase}
\affiliation{Department of Physics, Graduate School of Science, Kyoto University, Kyoto 606-8502, Japan}

\begin{abstract}
We explore how the quantum geometric properties of the Bloch wave function, characterized by the Hilbert-Schmidt quantum distance, impact magnetic phases in solid-state systems.
To this end, we investigate the spin susceptibility within the random phase approximation, considering the onsite Coulomb interaction.
We demonstrate that spin susceptibility can be decomposed into a trivial part, dependent solely on the band dispersion, and a geometric part, where the quantum distance plays a crucial role.
Focusing on a model of a quadratic band-touching semimetal, we show that a magnetic phase transition between ferromagnetic and antiferromagnetic order can be induced solely by tuning the wavefunction geometry, even while the energy spectrum is held constant. 
This highlights the versatility of quantum geometry as a mechanism for tuning magnetic properties independent of the energy spectrum.
Applying our framework to the Fe-pnictide and kagome lattice models, we further show that the geometric contribution is decisive in stabilizing their known antiferromagnetic and ferromagnetic states, respectively. 
Our work 
sheds light on the hidden quantum geometric aspects necessary
for understanding and engineering magnetic order in quantum materials.
\end{abstract}
\maketitle

\textit{Introduction---}
Quantum states are endowed with a geometric structure by defining an abstract distance, referred to as the Hilbert-Schmidt quantum distance, between wave functions~\cite{provost1980riemannian,shapere1989geometric}.
Mimicking the differential geometry, the quantum distance naturally gives rise to a metric concept known as the quantum geometric tensor, whose real and imaginary parts correspond to the quantum metric and Berry curvature, respectively~\cite{provost1980riemannian,shapere1989geometric,ma2010abelian}.
The Berry curvature, acting as an emergent magnetic field in solids, is responsible for the anomalous velocity of the carriers, leading to various Hall currents~\cite{xiao2010berry}.
Moreover, the Berry curvature is notable for its role in characterizing topological insulators and superconductors through the Chern number, a topological invariant obtained by integrating the Berry curvature over the Brillouin zone~\cite{haldane1988model,thouless1982quantized}.
Although the Berry curvature has been the subject of extensive research, other geometric quantities, such as quantum distance and quantum metric, have received relatively less attention. 
Recently, however, these geometric aspects have drawn increasing interest, as studies have begun to connect them to phenomena such as superfluid weight~\cite{peotta2015superfluidity,liang2017band,han2024quantum}, transport and optical properties~\cite{oh2025universal,oh2025color,ezawa2024analytic,das2023intrinsic,gao2023quantum,wu2024quantum,mitscherling2022bound,huhtinen2023conductivity,ghosh2024probing,neupert2013measuring,oh2024thermoelectric} including nonlinear Hall effect~\cite{gao2023quantum,wang2023quantum,han2024room}, Landau levels~\cite{rhim2020quantum,jung2024quantum,oh2024revisiting,hwang2021geometric}, bulk-edge correspondence~\cite{oh2022bulk,kim2023general},  
electron-phonon coupling~\cite{yu2024non}, and capacitance~\cite{komissarov2024quantum}.
Moreover, a few experimental results measuring these quantum geometric quantities have recently begun to emerge~\cite{gianfrate2020measurement,kang2024measurements,Tanaka2025,Banerjee2025}.

Recent studies have also explored connections between quantum geometry and various magnetic properties, including its relation to spin excitations in magnetic states~\cite{Wu2020,Herzog2022,Yu2024,Kang2024}, diamagnetism~\cite{oh2024revisiting}, ferromagnetic instability leading to spin triplet superconductivity~\cite{kitamura2024spin,Kitamura2025}, altermagnetic instability~\cite{Heinsdorf2024}, and odd-parity magnetic instability~\cite{Kudo2025}.
For instance, one study found that the geometry of the wavefunction can enhance diamagnetic susceptibility~\cite{oh2024revisiting}. 
The impact of quantum geometry on Landau levels suggests that the total energy in the presence of a magnetic field is intricately connected to quantum geometric factors, which, in turn, can drive a crossover between paramagnetism and diamagnetism. 
Another study has shown that quantum geometry can induce spin-triplet superconductivity, with the geometric structure of spin susceptibility favoring ferromagnetic fluctuations, thereby promoting spin-triplet pairing~\cite{kitamura2024spin}. 
While there is a growing body of research on how quantum geometry may influence magnetic properties, the potential for quantum geometry to induce magnetic phase transitions between ferromagnetic and antiferromagnetic states has remained largely unexplored.

In this paper, we investigate the nontrivial role of quantum geometry in the stabilization of ferromagnetic and antiferromagnetic states.
The key geometric quantity here is the Hilbert-Schmidt quantum distance, which quantifies the similarity between two quantum states as a dimensionless distance ranging from 0 to 1~\cite{buvzek1996quantum,dodonov2000hilbert}.
We begin by reformulating the general expression for spin susceptibility in a multiband system, separating it into a trivial component that depends solely on the band structure and a geometric component governed by the quantum distance.
To isolate these geometric effects, we employ a model of a quadratic band-touching semimetal, a system where the wavefunction geometry can be systematically modified while the energy spectrum remains invariant. In this model, we demonstrate that the overall geometric character is effectively controlled by a single parameter: the maximum value of the quantum distance over the Brillouin zone, denoted $d_\mathrm{max}$. This parameter serves as a direct measure of the interband mixing, allowing us to continuously tune the system from being geometrically trivial to nontrivial.
This allows us to show that a magnetic phase transition between ferromagnetic and antiferromagnetic ordering can be induced solely by varying the geometric character of the Bloch states. 
To connect our theory to real materials, we apply this geometric framework to Fe-pnictide and kagome lattice models, revealing that the geometric contribution to susceptibility is instrumental in stabilizing their respective antiferromagnetic and ferromagnetic orders. Our findings establish quantum geometry not merely as a perturbative influence but as a fundamental mechanism for controlling magnetic phases in solids.

\textit{Geometric contribution in spin susceptibility---}
To elucidate the role of quantum geometry, we analyze the static spin susceptibility, $\chi_{s}(\bm{q})$, of an interacting multiband system. We consider an on-site intra-orbital Coulomb interaction, \(U\), and treat it within the random phase approximation (RPA). In the simple single-band picture, the interacting susceptibility is given by
$\chi_{s}^{\mathrm{RPA}}(\bm{q})
  = \chi_{s}^{0}(\bm{q})/[1 - U\,\chi_{s}^{0}(\bm{q})]$.
This expression transparently shows that for a perturbative \(U\), a magnetic instability (the Stoner criterion) is first met at the wave vector \(\bm{q}\) where the bare susceptibility \(\chi_{s}^{0}(\bm{q})\) is maximal.
This core principle extends to the general multiband case. In the matrix formalism, the RPA susceptibility is
$\hat{\chi}_{s}^{\mathrm{RPA}}(\bm{q})
  = \hat{\chi}_{s}^{0}(\bm{q})
    \bigl[\hat{1} - \hat{U}\,\hat{\chi}_{s}^{0}(\bm{q})\bigr]^{-1}$,
where $\hat{U}$ is the matrix of Coulomb interactions in the orbital basis.
A magnetic instability occurs when the largest eigenvalue of the matrix product \(\hat{U}\,\hat{\chi}_{s}^{0}(\bm{q})\) approaches unity.
This instability condition is met at the wave vector $\bm{q}$ that maximizes this leading eigenvalue. For a simple interaction matrix proportional to the identity, this wave vector is determined by the peak of the largest eigenvalue of the bare susceptibility matrix, $\hat{\chi}^0_s(\bm{q})$. In many physical systems, the momentum-space structure of this largest eigenvalue is closely tracked by that of the total susceptibility, $\chi^0_s({\bm q})=\text{Tr}[\hat{\chi}^0_s(\bm{q})]$.
Therefore, in both single- and multi-band scenarios within the RPA framework, the magnetic ordering wave vector is primarily determined by the momentum-space structure of the bare susceptibility. This allows us to focus our analysis on the properties of \(\chi_{s}^{0}(\bm{q})\) to understand the intrinsic magnetic tendencies of the material.

Focusing on static fluctuations (see SI), the bare spin susceptibility can be written as
\begin{eqnarray}
\hspace{-3mm}
\chi^0_s(\bm{q})=2\sum_{nm}\int \frac{d\bm{k}}{(2\pi)^d}F_{nm}(\bm{k,q})[1-d^2_{nm}(\bm{k+q,k})], \label{eq:chi_bare_tot}
\end{eqnarray}
where $f(\epsilon) = (\exp[(\epsilon - \mu) / T] + 1)^{-1}$ with $\mu$ as the chemical potential, $F_{nm}(\bm{k,q})=[f(\epsilon_m(\bm{k}))-f(\epsilon_n(\bm{k+q}))]/[\epsilon_n(\bm{k+q})-\epsilon_m(\bm{k})]$, and $d^2_{nm}(\bm{k,k'})$ represents the quantum distance between $\ket{u_n(\bm{k})}$ and $\ket{u_m(\bm{k}')}$~\cite{buvzek1996quantum,dodonov2000hilbert,wilczek1989geometric}:
\begin{align}
   d_{nm}^2(\bm{k,k'}) =1-|\braket{u_n(\bm{k})|u_m(\bm{k}')}|^2.
\end{align}
Here, $\epsilon_n(\bm{k})$ and $\ket{u_n(\bm{k})}$ denote the $n$-th eigenvalue and eigenstate of $H_0(\bm{k})$, respectively. 
The formulation in Eq.~(\ref{eq:chi_bare_tot}) naturally invites a decomposition of the susceptibility into two parts. In a system with trivial geometry, the Bloch states
$\ket{u_n(\bm{k})}$ is independent of $\bm{k}$, leading to $d^2_{nm}(\bm{k, k'}) = 1 - \delta_{nm}$. Note that such a momentum-independent quantum distance arises in systems lacking inter-band coupling.
In this limit, only intraband ($n=m$) terms survive, and the susceptibility depends solely on the energy dispersion. This motivates the following separation~\cite{kitamura2024spin}:
\begin{eqnarray}
\chi^0_s=\chi^0_{s,\mathrm{band}}+\chi^0_{s,\mathrm{geom}}, \label{eq:suscept}
\end{eqnarray}
where the band susceptibility, $\chi^0_{s, \mathrm{band}}(\bm{q}) = 2 \sum_n \int \frac{d\bm{k}}{(2\pi)^d} F_{nn}(\bm{k, q})$, captures the physics determined solely by the band structure. The second term, the geometric susceptibility, $\chi^0_{s, \mathrm{geom}}(\bm{q}) = 2 \sum_{nm} \int \frac{d\bm{k}}{(2\pi)^d} F_{nm}(\bm{k, q}) [1 - d^2_{nm}(\bm{k+q, k}) - \delta_{nm}]$, contains all corrections arising from the nontrivial, momentum-dependent structure of the wavefunctions. 
By definition, $\chi^0_{s, \mathrm{geom}}(\bm{0}) = 0$,  meaning this term only influences magnetic ordering at finite wave vectors.
To ensure this decomposition is unambiguous, we establish a clear principle applied throughout this work: all quantities are defined strictly with respect to the bands of the primitive unit cell, which provides a unique and natural basis for our analysis.

The two components can favor different magnetic orderings. The band term, $\chi_\mathrm{s,band}^0$, may favor either ferromagnetic or antiferromagnetic fluctuations based on the band structure (e.g., density of states at the Fermi level or Fermi surface nesting between identical bands). 
Similarly, the geometric term, $\chi_\mathrm{s,geom}^0$,  which is driven by interband effects and wavefunction overlap, can also favor either ferromagnetic or antiferromagnetic correlations.
When these two contributions are in opposition, their competition can fundamentally alter the dominant magnetic instability of the system and even drive a phase transition, as we demonstrate later.

By examining the curvature of $\chi_s^0(\bm{q})$ at $\bm{q} = 0$, denoted as $\Omega_{\mu\nu} := \lim_{\bm{q} \to 0} \partial_{q_\mu} \partial_{q_\nu}\chi_s^0(\bm{q})$, we can establish a necessary condition for ferromagnetic ordering~\cite{kitamura2024spin}.
This condition arises from the requirement that ferromagnetic ordering necessitates a peak at $\bm{q} = 0$, i.e., $\mathrm{det}[\Omega] > 0$ with $\mathrm{Tr}[\Omega]<0$ in two-dimensional systems.
In contrast, if this necessary condition is not satisfied, $\chi_s^0(\bm{q})$ cannot have a peak at $\bm{q} = 0$. Given that the quantum metric arises from infinitesimal distances, the geometric contribution of $\Omega_{\mu\nu}$ inherently includes the quantum metric~\cite{kitamura2024spin} (see SI).

\textit{Application to a Quadratic Band Touching Model---}
%%%%%%%%%%%%%%%%%%%%%%%%%%%%%%%%%%%
\begin{figure}[t]
\includegraphics[width=85mm]{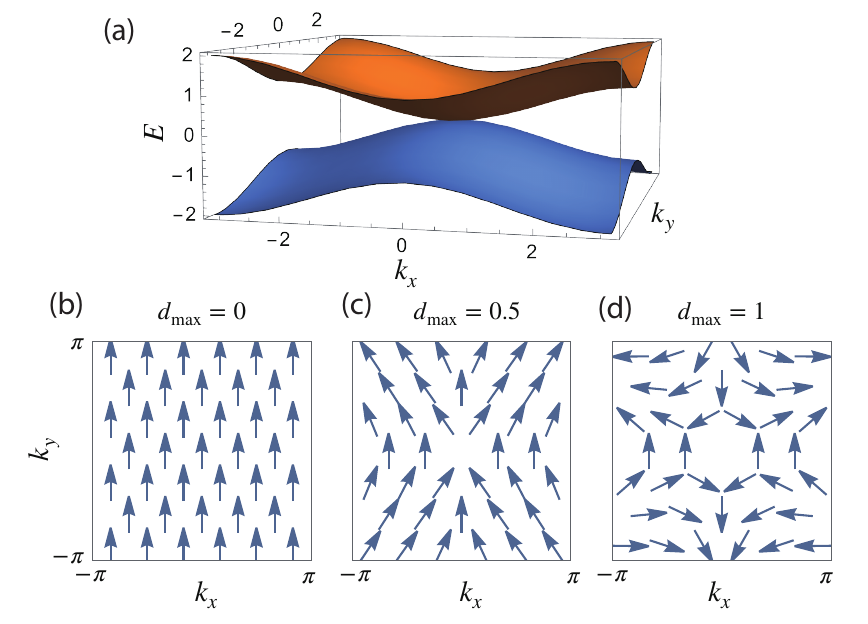} 
\caption{\label{fig1}
(a) Band dispersions in the quadratic band touching model, Eq.~(\ref{eq:energy}). (b-d) Pseudospin structures $(s_y(\bm{k}), s_z(\bm{k}))$ 
for (b) $d_\mathrm{max} =0$, (c) 0.5 and (d) 1. }
\end{figure}
%%%%%%%%%%%%%%%%%%%%%%%%%%%%%%%%%%%
To isolate the influence of quantum geometry from that of the energy spectrum, we now introduce a two-band model for a quadratic band touching semimetal described by the Hamiltonian:
    \begin{eqnarray}
        H_{0}({\bm{k}}) = \sum_{\alpha } h_\alpha ({\bm{k}}) \sigma_\alpha , \label{eq:Ham}
    \end{eqnarray}
where $\sigma_\alpha$ represents the identity matrix ($\alpha=0$) and Pauli matrices ($\alpha = x,y,z$). The components $h_{x,y,z} ({\bm{k}})$ are real quadratic functions defined as $h_{x} ({\bm{k}}) = d_\mathrm{max}\sqrt{1-d_\mathrm{max}^2}(1-\cos k_y)$, $h_{y} ({\bm{k}}) = 2d_\mathrm{max} \sin (k_x/2) \sin (k_y/2)$, $h_{z} ({\bm{k}}) = (1-d_\mathrm{max}^2)+\{(2 d_\mathrm{max}^2-1) \cos k_y-\cos k_x\}/2$, and $h_{0} ({\bm{k}}) = 0$. 
The parameter $d_\mathrm{max}$, which ranges from $0$ to $1$,  is defined as the maximum value of the quantum distance between $\ket{u_n(\bm{k})}$ and $\ket{u_n(\bm{k}')}$ over the Brillouin zone. The $d_\mathrm{max}$ remains the same for the upper ($n=+$) and lower ($n=-$) bands. The energy eigenvalues are given by 
\begin{align}
\epsilon_{\pm}(\bm{k})=\pm{\frac{2-\cos k_x -\cos k_y}{2}}, \label{eq:energy}
\end{align}
as illustrated in Fig.~\ref{fig1}(a). Notably, these eigenvalues are independent of $d_\mathrm{max}$.
This crucial feature allows us to vary the geometry of the wavefunctions while keeping the band structure fixed. The parameter $d_\mathrm{max}$ quantifies the maximum distinguishability between Bloch states within a band across the Brillouin zone (excluding the band touching point where the states are degenerate). As this intraband distance is a direct consequence of the interband coupling in the Hamiltonian, $d_\mathrm{max}$ acts as a single parameter that controls the overall geometric nontriviality of the bands, allowing us to tune the system from geometrically trivial ($d_\mathrm{max}=0$) to maximally nontrivial ($d_\mathrm{max}=1$).

In a two-band system, such geometric quantities are linked to a pseudospin $\bm{s}_{\bm{k}} =\bm{h}(\bm{k})/|\bm{h}(\bm{k})|$, with quantum distance and metric defined as $d^2_{nn}(\bm{k,k'})=(1-\bm{s}_{\bm{k}}\cdot \bm{s}_{\bm{k'}})/2$ and $g_{ij}^n(\bm{k})=\partial_{k_i}\bm{s}_{\bm{k}}\cdot \partial_{k_j}\bm{s}_{\bm{k}}/2$, respectively. 
Figures~\ref{fig1}(b–d) show pseudospin structures for $d_\mathrm{max}=0$, $0.5$, and $1$, highlighting the different geometries despite identical band structures.

%%%%%%%%%%%%%%%%%%%%%%%%%%%%%%%%%%%
\begin{figure}[t]
\includegraphics[width=85mm]{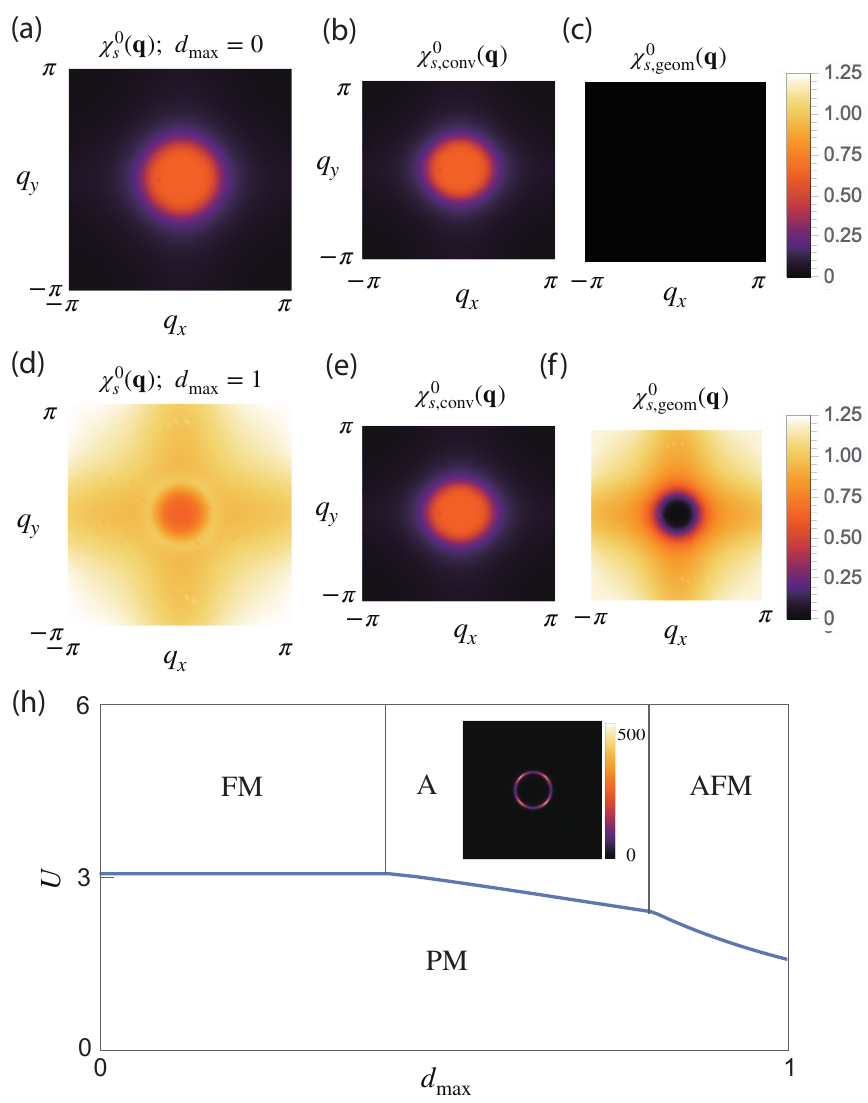} 
\caption{\label{fig2}
Noninteracting spin susceptibility in the quadratic band touching model in Eq.~(\ref{eq:Ham}) for (a-c) $d_\mathrm{max}=0$ and (d-f) $d_\mathrm{max}=1$. (a,d) $\chi^0_s(\bm{q})$, (b,e) $\chi^0_{s,\mathrm{band}}(\bm{q})$, and (c,f) $\chi^0_{s,\mathrm{geom}}(\bm{q})$. (h) Phase diagram between $d_\mathrm{max}$ and $U$ calculated by $\chi^{\mathrm{RPA}}$. In the calculations, we fix the parameters $T=0.01$ and $\mu =0.05$.}
\end{figure}
%%%%%%%%%%%%%%%%%%%%%%%%%%%%%%%%%%%

The impact of this purely geometric tuning is immediately evident in the noninteracting spin susceptibility.
We compute $\chi^0_s(\bm{q})$ for the two extremal cases: the geometrically trivial limit ($d_\mathrm{max}=0$)  and the maximally nontrivial limit ($d_\mathrm{max}=1$).
Figures~\ref{fig2}(a) and \ref{fig2}(d) show $\chi_s^0(\bm{q})$ for these two cases, respectively. 
As shown in Fig.~\ref{fig2}(a), the system at $d_\mathrm{max}=0$ exhibits a ferromagnetic fluctuation peaked at $\bm{q}=\bm{0}$.
Decomposing the susceptibility reveals that this arises entirely from the band contribution $\chi^0_{s,\text{band}}$, as the geometric part, $\chi^0_{s,\text{geom}}$, is identically zero [Figs.~\ref{fig2}(b,c)], consistent with the definition.

In stark contrast, the system at $d_\mathrm{max}=1$ displays a dominant antiferromagnetic fluctuation peaked at the zone corners~[Fig.~\ref{fig2}(d)].
Here, the band contribution remains unchanged and ferromagnetic, as shown in Fig.~\ref{fig2}(e).
However, the now-substantial geometric contribution, $\chi^0_{s,\text{geom}}$ introduces an antiferromagnetic tendency that overwhelms the band contribution, dictating the overall magnetic character of the system [Fig.~\ref{fig2}(f)].
This clearly demonstrates that a transition in magnetic tendency from ferromagnetic to antiferromagnetic can be driven solely by the quantum geometry of the wavefunctions.

To explore the impact of Coulomb interactions, we employ the RPA framework [details in SI]. The $d_\mathrm{max}$-$U$ phase diagram in Fig.~\ref{fig2}(h) shows the magnetic phases as a function of $d_\mathrm{max}$ and the Coulomb interaction strength $U$. 
For small $d_\mathrm{max}$, a ferromagnetic phase is stabilized above a critical interaction $U^*$. This critical value is constant in the ferromagnetic regime because 
{$d_{nm}(\bm{k},\bm{k})=1-\delta_{nm}$ for any ${\bm k}$ leads to $\chi^0_s(\bm{0}) = \chi^0_{s,\text{band}}(\bm 0)$, making the critical value independent of $d_\mathrm{max}$.}
In contrast, for large $d_\mathrm{max}$, an antiferromagnetic phase emerges.
Here, the critical $U^*$ for the antiferromagnetic transition does depend on $d_\mathrm{max}$, as intensity of $\chi^0_{s,\mathrm{geom}}(\pm\pi,\pm\pi)$ is controlled by the geometry. 
Between the ferromagnetic and antiferromagnetic phases, a different phase (labeled `A') emerges, marked by peak positions distinct from those of the ferromagnetic and antiferromagnetic orders.
This suggests that quantum geometry can serve as a rich tuning parameter, potentially giving rise to more complex magnetic states like incommensurate spin density waves.

\textit{Geometry-enabled Antiferromagnetic Ordering---}
Having established a model where geometry drives a magnetic transition, we now apply our framework to a real material system: the Fe-pnictides. These materials are a canonical example of itinerant magnetism, where antiferromagnetic fluctuations are believed to arise from Fermi surface nesting between disconnected electron and hole pockets~\cite{Kuroki2008}. Here, we demonstrate that this nesting-driven magnetism can be reinterpreted as a direct consequence of the system's quantum geometry.

\begin{figure}[t]
\includegraphics[width=85mm]{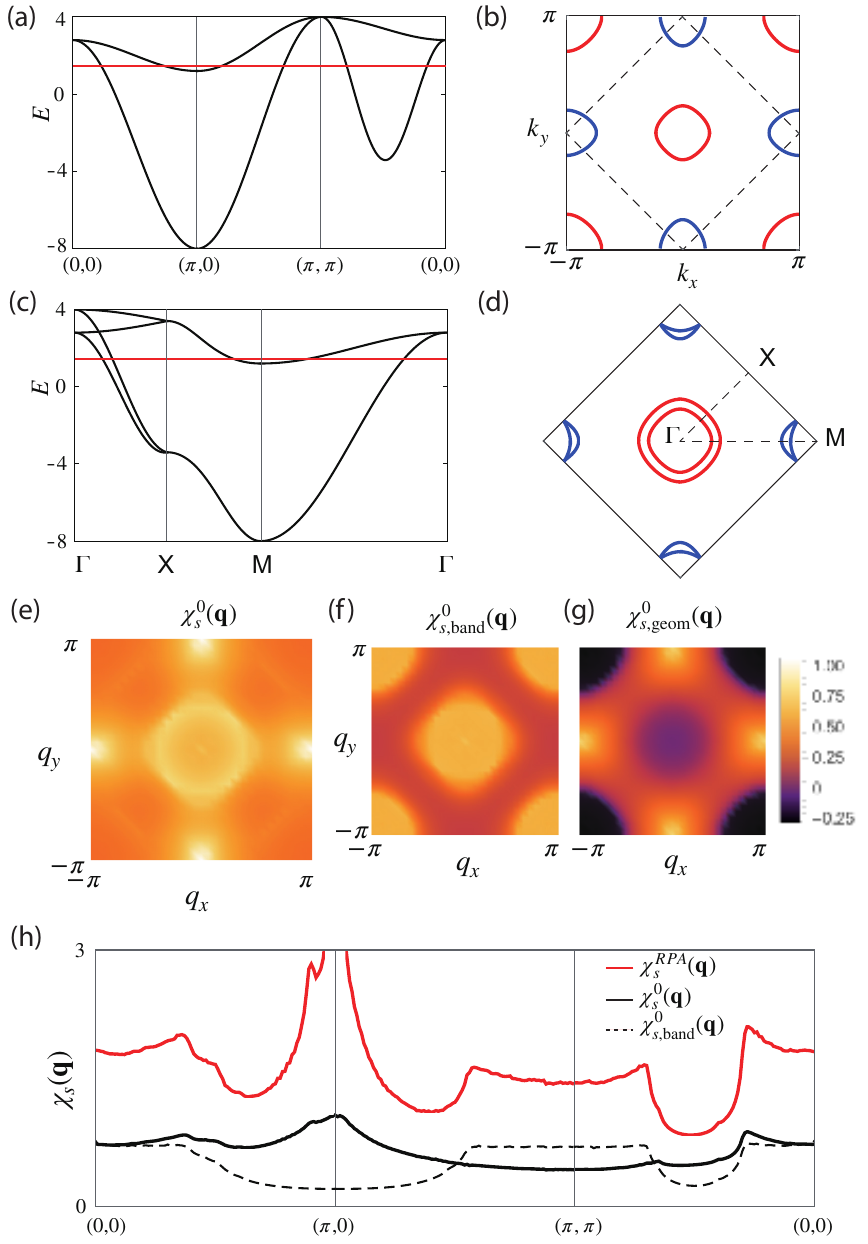} 
\caption{\label{fig3}
(a) The band structure of the two-band model {in Ref.~\cite{raghu2008minimal}.} (b) The Fermi surface at $\mu=1.45$ on the 1Fe/Cell BZ. The dashed lines indicate the folded BZ for the 2Fe/Cell. (c) The band structure folded to the small BZ. (d) The Fermi surface folded down into the the 2Fe/Cell BZ. The red lines in (a) and (c) represent the chemical position $\mu = 1.45$. The blue and red curves in (b) and (d) represent hole and electron Fermi pockets. (e-g) The noninteracting spin susceptibility; (e) the total susceptibility $\chi^0_s (\bm{q})$, (f) band contribution $\chi^0_{s,\mathrm{band}}(\bm{q})$ and (g) geometric contribution $\chi^0_{s,\mathrm{geom}}(\bm{q})$. (h) %Spin suscetibilities. 
The red solid line shows the spin susceptibility obtained by the RPA, $\chi^{\rm RPA}_s (\bm{q})$.  
We show $\chi^0_s (\bm{q})$ (black solid line) and $\chi^0_{s,\mathrm{band}}(\bm{q})$ (black dashed line) for comparison. In the calculations, we set the parameters $T=0.01$, $U=1.7$, and $\mu =1.45$.}
\end{figure}

We employ the minimal two-band model for Fe-pnictides~\cite{raghu2008minimal}, which effectively captures the essential features of the band structure and orbital hybridization responsible for magnetic fluctuations in this class of materials. The detailed form of the tight-binding Hamiltonian and its parameters are provided in the SI.
The model's band structure and Fermi surface are shown in both the unfolded (primitive 1-Fe/cell) and folded (2-Fe/cell) Brillouin zones [Figs.~\ref{fig3}(a-d)]. 
Presenting both zones highlights a crucial aspect of our method: since Brillouin zone folding can ambiguously reassign physical processes between the intraband and interband contributions, we perform our decomposition strictly within the primitive (unfolded) zone to ensure a unique and physically meaningful analysis.
The folded representation is included for comparison with convention and first-principle results~\cite{PhysRevLett.100.237003,PhysRevLett.101.057003,hirschfeld2011gap}.

The concept of nesting between disconnected Fermi surfaces is fundamentally a statement about the wavefunction overlap between states on these surfaces. 
 A finite overlap is required for interband scattering to contribute to spin fluctuations; without it, nesting would only lead to higher-order multipole fluctuations~\cite{chubukov2008magnetism,Kuroki2008,shimizu2018two}. This crucial overlap is precisely what is quantified by the quantum distance in our formalism. Therefore, the geometric susceptibility, $\chi^0_{s,\text{geom}}$, provides a natural and quantitative framework for understanding nesting-driven magnetism.

Our numerical calculations confirm this picture. The total noninteracting susceptibility, $\chi^0_s$, correctly reproduces the well-known antiferromagnetic fluctuations peaked at $(\pm\pi,0)$ and $(0,\pm \pi)$ [Fig.~\ref{fig3}(e)]. Decomposing the susceptibility reveals the origin of this behavior. 
The band contribution, $\chi^0_{s,\text{band}}$, shows only broad, weak features and does not favor the antiferromagnetic ordering [Fig.~\ref{fig3}(f)]. In contrast, the geometric contribution, $\chi^0_{s,\text{geom}}$, exhibits sharp peaks exactly at the antiferromagnetic wave vectors [Fig.~\ref{fig3}(g)]. It is the geometric term, driven by the interband overlap (i.e., nesting) between the electron and hole pockets, that entirely determines the magnetic character of the system.
The inclusion of an interaction $U$ within the RPA framework then stabilizes this antiferromagnetic ordering, as shown in Fig.~\ref{fig3}(h). This example illustrates how our geometric decomposition provides a clear and fundamental interpretation of magnetism in a real material.

%%%%%%%%%%%%%%%%%%%%%%%%%%%%%%%%%%%
%%%%%%%%%%%%%%%%%%%%%%%%%%%%%%%%%%%

\textit{Geometry-Induced Ferromagnetic Ordering---}
The preceding examples have shown the geometric contribution leading to antiferromagnetic correlations. However, the role of quantum geometry is not limited to a single type of ordering; it can also be the deciding factor in promoting ferromagnetism. To illustrate this, we now examine the kagome lattice model, which is known to host a ferromagnetic instability near the van Hove singularity filling ($n=5/12$)~\cite{kiesel2013unconventional}. While the large density of states at the Fermi level provides a general tendency towards magnetism, the selection of the ferromagnetic channel is a more subtle effect rooted in the lattice structure.

Indeed, it has been previously established that in frustrated itinerant systems like the kagome lattice, ferromagnetism can be stabilized through a mechanism known as sublattice interference~\cite{martin2008itinerant, kiesel2013unconventional}. 
The key idea is that magnetic susceptibilities at nonzero ordering wave vectors, such as those associated with staggered magnetism at the M points, require coherent contributions from different sublattices. However, the Bloch wavefunctions on the kagome lattice carry nontrivial relative phases between sublattice components. When evaluating particle–hole scattering amplitudes at these wave vectors, the nontrivial relative phases cause destructive interference between contributions from different sublattices, thereby strongly suppressing the corresponding antiferromagnetic fluctuations. By contrast, the uniform ${\bm q}=0$ channel, which does not rely on inter-sublattice coherence, is unaffected by this cancellation. As a result, the ferromagnetic instability remains the dominant ordering tendency.

\begin{figure}[t]
\includegraphics[width=85mm]{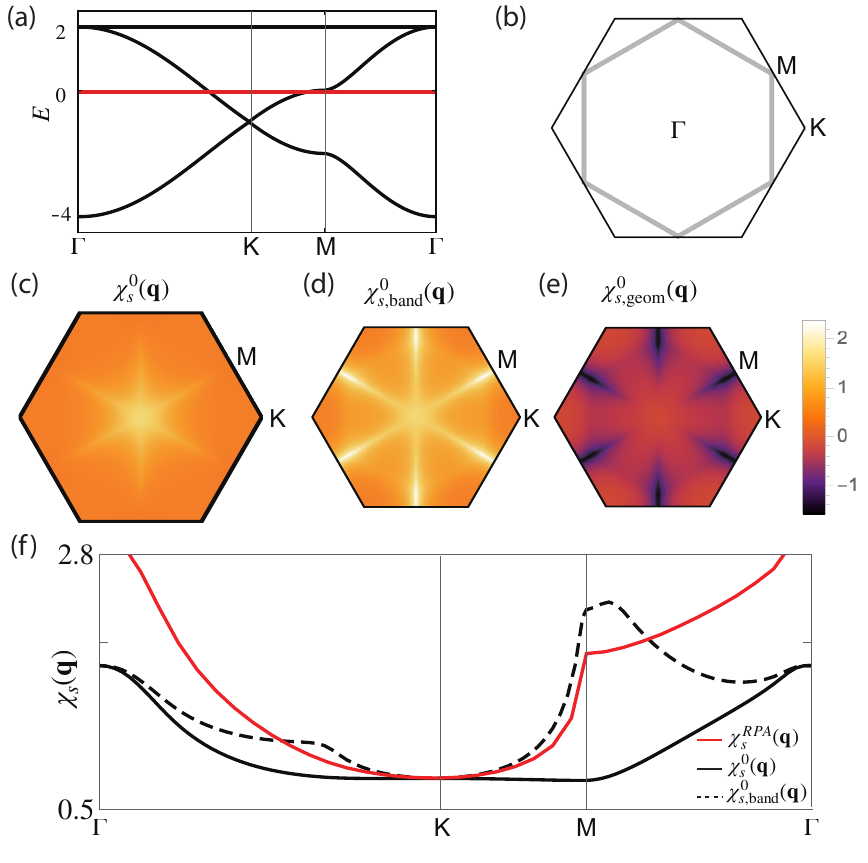} 
\caption{\label{fig4}
(a) Energy band dispersion in the kagome lattice. The red line represents the chemical potential $\mu=-0.01$. (b) Fermi surface at a van Hove filling $\mu=0$. (c-e) The noninteracting spin susceptibility; (c) the total susceptibility $\chi^0_s (\bm{q})$, (d) conventional contribution $\chi^0_{s,\mathrm{band}}(\bm{q})$ and (e) geometric contribution $\chi^0_{s,\mathrm{geom}}(\bm{q})$. (f) 
The black solid and dashed lines represent  $\chi^0_s (\bm{q})$ and $\chi^0_{s,\mathrm{band}}(\bm{q})$, respectively. The red solid line represents $\chi^{\rm RPA}_s (\bm{q})$. In the calculations, we fix the parameters $T=0.01$, $U=1$, and $\mu =-0.01$.}
\end{figure}

Our geometric decomposition provides a clear and quantitative framework for understanding this phenomenon. As shown in the band structure and Fermi surface plots [Figs.~\ref{fig4}(a,b)], the system is poised for instability. The band contribution to the susceptibility, $\chi^0_{s,\text{band}}$, reflects the high density of states by showing strong magnetic fluctuations at multiple wave vectors, including both the $\Gamma$ point ($\bm q=\bm 0$) and the M points [Fig.~\ref{fig4}(d)]. This term, being blind to the wavefunction's internal structure, suggests a competition between ferromagnetic and staggered magnetic orderings.
The geometric contribution, $\chi^0_{s,\text{geom}}$, however, fundamentally alters this picture by precisely capturing the effect of sublattice interference [Fig.~\ref{fig4}(e)]. Unlike in the Fe-pnictide case, here the geometric term is strongly negative at the M points, while being zero at the $\Gamma$ point by definition. This negative contribution directly quantifies the destructive interference for the staggered magnetic susceptibility. Consequently, the geometric term selectively cancels the magnetic fluctuations at the M points, leaving the ferromagnetic fluctuation at $\bm q=\bm 0$ as the dominant instability in the total susceptibility, $\chi^0_s$ [Fig.~\ref{fig4}(c)].
The inclusion of interactions via the RPA then naturally leads to a stable ferromagnetic state [Fig.~\ref{fig4}(f)]. This example powerfully demonstrates how our formalism connects the abstract concept of quantum geometry to the well-established physical mechanism of sublattice interference, clarifying how geometry can determine the magnetic ground state by suppressing competing orders.

%%%%%%%%%%%%%%%%%%%%%%%%%%%%%%%%%%%
%%%%%%%%%%%%%%%%%%%%%%%%%%%%%%%%%%%
\textit{Conclusion---}
In this study, we have established that the quantum geometry of Bloch wavefunctions can act as a primary driving mechanism for magnetic phase transitions in solids. By reformulating the spin susceptibility we separated the contributions from the energy dispersion and wavefunction geometry, with the latter being governed by the Hilbert-Schmidt quantum distance.

Our analysis of a quadratic band-touching model, where the geometry could be tuned independently of a fixed energy spectrum, provided a clear demonstration of this principle: a transition from a ferromagnetic to an antiferromagnetic instability was induced solely by varying the geometric character of the quantum states. This finding highlights a new pathway for engineering magnetic properties that is decoupled from the band structure.

Furthermore, we demonstrated the power of this framework by reinterpreting the magnetism in well-known material classes. For Fe-pnictides, we showed that the geometric susceptibility, $\chi^0_{s,\text{geom}}$, provides a quantitative and fundamental basis for the established mechanism of Fermi surface nesting. Conversely, for the kagome lattice, we found that $\chi^0_{s,\text{geom}}$ actively promotes ferromagnetism by suppressing competing staggered fluctuations via sublattice interference. These examples confirm that quantum geometry can decisively determine the magnetic ground state, either by enhancing a specific ordering channel or by suppressing others.

Our work opens several avenues for future research. The geometric framework presented here can be used to reinterpret the origins of magnetism in a wide range of existing materials. Theoretically, it can be extended to investigate more complex phenomena, such as the geometry's role in stabilizing incommensurate or non-collinear magnetic orders. Computationally, integrating our approach with first-principles calculations could help predict novel materials where geometry-driven magnetic responses are prominent. Finally, our findings suggest the exciting possibility of dynamically controlling magnetic phases through external knobs such as strain or electric fields that directly manipulate the quantum geometry of wavefunctions, paving the way for new applications in quantum spintronics.
\begin{acknowledgments}
The authors thank M. Shimizu for useful discussions. C.O. was supported by Q-STEP, WINGS Program, the University of Tokyo. T. Kitamura, A. Daido, and Y. Yanase were supported by JSPS KAKENHI (Grants Nos. JP21K13880, 
JP21J14804, JP22H04476, JP22H04933,
JP22H01181, JP22J22520, JP23K17353, JP23K22452, JP24H00007,
JP24K21530, JP25H01249), JSPS research fellowship, and WISE Program
MEXT.
J.W.R was supported by the National Research Foundation of Korea (NRF) Grant funded by the Korean government (MSIT) (Grant nos. 2021R1A2C1010572 and 2022M3H3A1063074) and the Ministry of Education (Grant no. RS-2023-00285390).
\end{acknowledgments}

\bibliography{ref.bib}

\onecolumngrid

\clearpage

\appendix

\section{Noninteracting spin susceptibility}
We start from the Matsubara expression for the noninteracting static spin susceptibility
\begin{equation}
\chi_s^{0}(\bm q,i\Omega_m)
=\frac{-2}{V\beta}\sum_{\bm k,\omega_n}\mathrm{Tr}\bigl[G(\bm{k+q},i\omega_n+i\Omega_m)\,G(\bm k,i\omega_n)\bigr],
\end{equation}
with $G(\bm k,i\omega_n)=[i\omega_n-H_0(\bm k)]^{-1}$. Working in the band eigenbasis of $H_0(\bm k)$ the Green's function is diagonal in band indices, $G_{mm'}(\bm k,i\omega_n)=\delta_{mm'}(i\omega_n-\epsilon_m(\bm k))^{-1}$. Evaluating the trace in the band basis yields
\begin{equation}
\mathrm{Tr}\bigl[G(\bm{k+q},i\omega_n+i\Omega_m)G(\bm k,i\omega_n)\bigr]
=\sum_{n,m}\frac{\bigl|\braket{u_n(\bm{k+q})|u_m(\bm k)}\bigr|^2}
{(i\omega_n+i\Omega_m-\epsilon_n(\bm{k+q}))(i\omega_n-\epsilon_m(\bm k))}.
\end{equation}

The standard Matsubara summation identity is
\begin{equation}
\frac{1}{\beta}\sum_{\omega_n}\frac{1}{(i\omega_n-a)(i\omega_n+i\Omega_m-b)}
=\frac{f(a)-f(b)}{i\Omega_m+a-b}.
\end{equation}
Setting $i\Omega_m\to 0$ for the static limit gives
\begin{equation}
\frac{1}{\beta}\sum_{\omega_n}\frac{1}{(i\omega_n-\epsilon_m)(i\omega_n-\epsilon_n)}
=\frac{f(\epsilon_m)-f(\epsilon_n)}{\epsilon_n-\epsilon_m}.
\end{equation}
Applying this to the trace and absorbing the volume factor into the momentum integral leads to
\begin{equation}
\chi_s^{0}(\mathbf{q}) = 2\sum_{n,m}\int \frac{d\mathbf{k}}{(2\pi)^d}
F_{nm}(\mathbf{k},\mathbf{q})
\bigl|\langle u_n(\mathbf{k+q})|u_m(\mathbf{k})\rangle\bigr|^2,
\end{equation}
which can be recast using $d_{nm}^2 = 1 - |\langle u_n(\mathbf{k+q})|u_m(\mathbf{k})\rangle|^2$ as in the main text.
Separating the intraband part yields the decomposition into $\chi_{s,\mathrm{band}}^{0}$ and $\chi_{s,\mathrm{geom}}^{0}$.  

\section{Random Phase Approximation}

\subsection{General Formulation of RPA Spin Susceptibility}

In the random phase approximation (RPA), the interacting spin susceptibility is obtained by summing bubble diagrams to infinite order. For multiband systems, the spin susceptibility becomes a matrix with respect to orbital indices, and the RPA spin susceptibility is given by
\begin{equation}
    \hat{\chi}^{\mathrm{RPA}}_s(\bm{q},\omega) = 
    \left[\hat{1} - \hat{U}^s \hat{\chi}^0(\bm{q},\omega)\right]^{-1} 
    \hat{\chi}^0(\bm{q},\omega),
\end{equation}
where \( \hat{\chi}^0(\bm{q},\omega) \) is the irreducible (bare) spin susceptibility matrix and \( \hat{U}^s \) is the interaction matrix. We only consider onsite Coulomb interaction \( \hat{U}^s=U\delta_{l_1,l_2}\delta_{l_2,l_3}\delta_{l_3,l_4} \). The orbital-resolved components of the bare susceptibility are given by
\begin{align}
    [\chi^0(\bm{q}, i\omega_n)]^{\ell_1 \ell_2 ; \ell_3 \ell_4} = 
    -\frac{1}{N} \sum_{\bm{k}, \mu, \nu}
    a^{\ell_1}_{\mu}(\bm{k}) a^{\ell_3 *}_{\mu}(\bm{k})
    a^{\ell_4}_{\nu}(\bm{k}+\bm{q}) a^{\ell_2 *}_{\nu}(\bm{k}+\bm{q})
    \frac{f(\epsilon_{\bm{k} \mu}) - f(\epsilon_{\bm{k}+\bm{q} \nu})}
         {i\omega_n + \epsilon_{\bm{k} \mu} - \epsilon_{\bm{k}+\bm{q} \nu}},
\end{align}
where \( a^{\ell}_{\mu}(\bm{k}) \) is the amplitude of orbital \( \ell \) in band \( \mu \) at momentum \( \bm{k} \), and \( f(\epsilon) \) is the Fermi-Dirac distribution function.

In the single-band limit, this reduces to a scalar expression:
\begin{equation}
    \chi_s^{\mathrm{RPA}}(\bm{q},\omega) = 
    \frac{\chi^0(\bm{q},\omega)}{1 - U \chi^0(\bm{q},\omega)}.
\end{equation}

\subsection{Choice of RPA Scheme}

It is important to choose the appropriate RPA formulation based on the nature of the model under consideration. In the case of the \( d_{\mathrm{max}} \) model, although the Hamiltonian has a two-band structure, the model is intentionally designed as a minimal platform to isolate the effect of quantum geometry—specifically, the quantum distance—on magnetic fluctuations. Notably, while the energy spectrum remains independent of the parameter \( d_{\mathrm{max}} \), the wavefunction structure varies with it.

%\TK{In the full RPA scheme,  $d_{\rm {max}}$ plays two different roles. First, it changes $\chi_{0}(\bm q)$ as shown in the main text. Second, it modifies the ratio of the mixing between different magnetic multipole orders. Therefore, to focus on the effect of $d_{\rm {max}}$ on $\chi_{0}(\bm q)$, we adopt scalar expression of mutli-band stoner criteria given by $\tilde{U}\chi_0(\bm q)\geq1$~\cite{Nogaki2024}, where mixing of different multipole order is ignored. Here, $\tilde{U}$ is the renomalized onsite Coulomb interaction and $\tilde{U} = U/2$ is satisfied in $d_{\mathrm{max}}$ model. This approach allows us to maintain a clear focus on the influence of quantum geometry on $\chi_0(\bm q)$.}

If the full multiorbital RPA formalism is applied to the \( d_{\mathrm{max}} \) model, the momentum-space peak of the RPA spin susceptibility \( \chi^\mathrm{RPA}_s(\bm{q}) \) may deviate from that of the bare susceptibility \( \chi^0_s(\bm{q}) \) when the interaction strength \( U \) is large. In particular, the Stoner instability condition, which requires the maximum eigenvalue of the Stoner matrix to reach unity, may be satisfied at a momentum different from that favored by the bare susceptibility. However, such large values of \( U \) lie outside the perturbative regime, where the RPA approximation is no longer quantitatively reliable.
Given that RPA is fundamentally a perturbative method, its results are physically meaningful only when \( U \) is sufficiently small. For this reason, we adopt the scalar form of RPA for the \( d_{\mathrm{max}} \) model to avoid unphysical artifacts. This approach allows us to maintain a clear focus on the influence of quantum geometry on magnetic response, without complications introduced by nonperturbative effects.

Furthermore, our simplified approach is validated by recent work that formally derives a scalar Stoner criterion for multiband systems under the approximation of neglecting multipole mixing~\cite{Nogaki2024}. This leads to a condition of the form $\tilde{U}\chi^0_s(\bm{q}) \geq 1$, where $\tilde{U}$ is a renormalized interaction. While this implies that the critical interaction strength in our phase diagram might be quantitatively underestimated (e.g., $\tilde{U}=U/2$ for $d_\mathrm{max}=1$), it validates our core assumption that the magnetic instability is governed by the peaks of the total bare susceptibility $\chi^0_s(\bm{q})$. This reinforces the physical picture presented in the main text.

By contrast, for more realistic systems with intrinsic multiorbital character—such as the kagome lattice model and the Fe-pnictide model—we employ the full multiorbital RPA scheme, which correctly incorporates the orbital degrees of freedom and their interaction-driven feedback. In these cases, we have verified that the peak positions of the spin susceptibility remain consistent even when the scalar RPA approximation is used. This consistency reinforces the robustness of the physical picture obtained.

\section{Quantum metric and distance}
The HS distance between Bloch states at $\bm k$ and $\bm k+d \bm k$ is 
\begin{align}
    d^2_{nn}\left(\bm k,\bm k+ d\bm k\right) & = 1- |\braket{u_n(\bm{k})|u_n(\bm{k}+d\bm{k}})|^2, \\
    &= 1-  \braket{u_n(\bm{k})|u_n(\bm{k}+d\bm{k})}\braket{u_n(\bm{k}+d\bm{k})|u_n(\bm{k})}, \\
    &\simeq 1-   \braket{u_n(\bm{k})|(|u_n(\bm{k})}+\sum_\mu\ket{\partial_{k_\mu}u_n(\bm{k})}dk_\mu)(\bra{u_n(\bm{k})}+\sum_\nu\bra{\partial_{k_\nu}u_n(\bm{k})}dk_\nu)\ket{u_n(\bm{k})}, \\
    &=  \sum_{\mu,\nu} g_n^{\mu\nu} dk_\mu dk_\nu, 
\end{align}
where $g^{\mu\nu}_n(\bm{k})=\sum_{m(\neq n)}A^\mu_{nm}(\bm{k})A^\nu_{mn}(\bm{k})+ c.c.$ is the quantum metric, with
$A^\mu_{nm}(\bm{k})= -i \braket{u_n(\bm{k})|\partial_{k_\mu}|u_m(\bm{k})}$.

For two band systems, the HS distance can be written with pseudospins.  The inner product is written as 
\begin{align}
    |\braket{u_n(\bm{k})|u_n(\bm{k'})}|^2=\braket{u_+(\bm{k'})|\frac{1}{2}+\frac{1}{2}\bm{s}_{\bm{k}}\cdot\bm{\sigma}|u_+(\bm{k'})}=\frac{1}{2}+\frac{1}{2}\bm{s}_{\bm{k'}}\cdot \bm{s}_{\bm{k}},
\end{align}
where $s_{\bm{k}}=\braket{u_+(\bm{k})|\bm{\sigma}|u_+(\bm{k})}$ is pseudospin for upper band.
Thus, $d^2_{nn}(\bm{k,k'})=(1-\bm{s}_{\bm{k}}\cdot \bm{s}_{\bm{k'}})/2$.

\section{Necessary condition for ferromagnetic ordering}
By considering the curvature of $\chi_s^0$ at $\bm{q}=0$, one can obtain the necessary condition for ferromagnetic ordering
%or sufficient condition for antiferromagnetic fluctuation 
\cite{kitamura2024spin}.
This condition comes from the fact that the ferromagnetic ordering requires the peak at $\bm{q}=0$, i.e., $\Omega_{\mu\nu} :=\lim_{\bm{q}\to 0} \partial_{q_\mu} \partial_{q_\nu} \chi_s^0(\bm{q}) < 0$. If $\Omega_{\mu\nu} > 0$, peaks of $\chi_s^0(\bm{q})$ always exist at $\bm{q\neq0}$. 
%Furthermore, this quantity $\chi_s^0/2$ has the meaning of generalized electric susceptibility, which generalizes the concept of electric susceptibility from insulators to metals \cite{kitamura2024spin,shitade2018theory,shitade2019theory,gao2018microscopic}.
Similar to the susceptibility $\chi_s^0$, the curvature $\Omega_{\mu\nu}$ can be seperated into two terms \cite{kitamura2024spin}:
\begin{eqnarray}
\Omega_{\mu\nu}=\Omega_{\mu\nu,\mathrm{band}}+\Omega_{\mu\nu,\mathrm{geom}},
\end{eqnarray}
where 
\begin{align}
    &\Omega_{\mu\nu,\mathrm{band}}=-4\sum_n \int \frac{d^2k}{(2\pi)^2}\left(\frac{f^{(2)}(\epsilon_n(\bm{k}))}{12}[M_n^{-1}]_{\mu\nu}\right),\\
    &\Omega_{\mu\nu,\mathrm{geom}}= \nonumber \\
    &4\sum_n\int \frac{d^2k}{(2\pi)^2}\left( \frac{f'(\epsilon_n(\bm{k}))}{2}g^{\mu\nu}_n(\bm{k})+f(\epsilon_n(\bm{k}))X^{\mu\nu}_n(\bm{k})\right),
\end{align}
where $f'=\partial_\epsilon f$, $f^{(2)}=\partial_\epsilon \partial_\epsilon f$, $[M^{-1}_n]_{\mu\nu}=\partial_{k_\mu}\partial_{k_\nu}\epsilon_n(\bm{k})$ is the inverse mass tensor of $n$-th band, $g^{\mu\nu}_n(\bm{k})=\sum_{m(\neq n)}A^\mu_{nm}(\bm{k})A^\nu_{mn}(\bm{k})+ c.c.$ is the quantum metric, which is the real part of the quantum geometric tensor, and $X^{\mu\nu}_n =\sum_{m(\neq n)}(A^\mu_{nm}(\bm{k})A^\nu_{mn}(\bm{k})+ c.c)/(\epsilon_m(\bm{k})-\epsilon_n(\bm{k}))$ is the positional shift \cite{GaoYangPRL}. $A^\mu_{nm}(\bm{k})= -i \braket{u_n(\bm{k})|\partial_{k_\mu}|u_m(\bm{k})}$ is the interband Berry connection. 
The first term in $\Omega_{\mu\nu,\mathrm{geom}}$ contributes negatively, while the second term contributes positively. Thus, the sign of $\Omega_{\mu\nu,\mathrm{geom}}$ is determined by which of the first and second terms contributes significantly.

Similar to $\chi^0$, the conventional term $\Omega_{\mu\nu,\mathrm{band}}$ and geometric term $\Omega_{\mu\nu,\mathrm{geom}}$ can also be positive or negative. 
When $\Omega\mathrm{s,band}^0$ and $\Omega\mathrm{s,geom}^0$ contribute oppositely, they can contribute to a form of $\chi_{s}^0$, potentially inducing a magnetic phase transition driven by quantum geometry.

\section{Geometric tensor of quadratic band touching model}
The quantum metric of the model in eq.~(5) of the main text is given by 
\begin{align}
    &g^n_{xx}(\bm{k})=d_\mathrm{max}^2\frac{2 \cos^2 (k_x/2) \sin^2 (k_y/2)}{(-2+\cos k_x+\cos k_y)^2},  \nonumber \\
    &g^n_{yy}(\bm{k})=d_\mathrm{max}^2\frac{2 \cos^2 (k_y/2) \sin^2 (k_x/2)}{(-2+\cos k_x+\cos k_y)^2},  \nonumber \\
    &g^n_{xy}(\bm{k})=-d_\mathrm{max}^2\frac{ \sin (k_x) \sin (k_y)}{2(-2+\cos k_x+\cos k_y)^2}.
\end{align}
At the same time, the Berry curvature is zero.

\section{Tight-Binding Model for Fe-Pnictides}
In the main text, we analyze the magnetic properties of Fe-pnictides using a minimal two-band model from Ref.~\cite{raghu2008minimal}. The tight-binding Hamiltonian is given by: \begin{align} H_{Fe}(\bm{k})= \epsilon_+(\bm{k})\sigma_0+\epsilon_-(\bm{k})\sigma_3+\epsilon_{xy}(\bm{k})\sigma_1, \end{align} with its components defined as: 
\begin{align}
    &\epsilon_{\pm}(\bm{k}) = \frac{\epsilon_x(\bm{k})\pm \epsilon_y(\bm{k})}{2},\nonumber\\
    &\epsilon_x(\bm{k}) = -2t_1 \cos k_x - 2t_2 \cos k_y - 4t_3 \cos k_x \cos k_y,\nonumber\\
    &\epsilon_y(\bm{k}) = -2t_2 \cos k_x - 2t_1 \cos k_y - 4t_3 \cos k_x \cos k_y, \nonumber\\
    &\epsilon_{xy}(\bm{k}) = -4t_4 \sin k_x \sin k_y.
\end{align}
For our calculations, we adopt the hopping parameters $(t_1, t_2, t_3, t_4)$ from \cite{raghu2008minimal}.
\end{document}